# Ferromagnetism and structural phase transition in monoclinic FeGe film

Guangdong Nie[1,2], Guanghui Han[3], Erfa S. Z.[1,2], Kangxi Liu[1], Shijian Chen[1], Hao Ding[1,2], Fangdong Tang[1,2], Licong Peng[3], Young Sun[1,2] & Deshun Hong[1,2]*

[1]Department of Applied Physics, Chongqing University, Chongqing 401331, China

[2]Center of Quantum Materials and Devices, Chongqing University, Chongqing 401331, China

[3] School of Materials Science and Engineering, Peking University, Beijing 100871, China

**Abstract:**

Binary compound FeGe hosts multiple structures, from cubic and hexagonal to monoclinic. Compared to the well-known skyrmion lattice in the cubic phase and the antiferromagnetic charge-density wave in the hexagonal phase, the monoclinic FeGe is less explored. Here, we synthesized the monoclinic FeGe films on $Al_2O_3$ (001) and studied their structural, magnetic, and transport properties. X-ray diffraction and transmission electron microscopy characterizations indicate that the FeGe films are epitaxial to the substrate. Unlike the antiferromagnetic bulk, the monoclinic FeGe films are ferromagnetic with Curie-temperature as high as ~ 800 K, contributing to the anomalous Hall effect in the transport measurements. Similar to the hexagonal FeGe, we captured a structural phase transition in the monoclinic FeGe films at ~ 100 K in real and reciprocal spaces by transmission electron microscope. Our work enriches the phase diagram of the FeGe family and suggests that FeGe offers an ideal platform for studying multiphase transitions and related device applications.

Physical properties and structures are strongly intertwined, thus a material hosting various structures is of great interest. Binary compound FeGe is one of these materials, which can crystalize into cubic, hexagonal, and monoclinic structures, as shown in Fig. 1 (a). The cubic FeGe, also known as chiral B20, is helimagnetic and famous for skyrmion lattice due to the inversion symmetry breaking induced Dzyaloshinskii-Moriya interaction (DMI) [1-3]. In epitaxial films, its skyrmion phases can extend temperature and field ranges [1, 4]. The hexagonal FeGe has recently become a research spotlight whose lattice is formed by alternative stacking of $Fe_3Ge$ kagome layers and Ge honeycomb layers [5-7]. This unique lattice geometry can induce electronic and magnetic frustrations. As a consequence, rich physics such as flat band, massless Dirac dispersion, and van

---

*  dhong@cqu.edu.cn

Hove singularities can emerge [8-11], which offers an ideal platform for fractional excitations [12-13], superconductivity [14] and quantum spin liquid [15]. Moreover, a dimerization of Ge along the *c*-axis induced charge density wave (CDW) below ~ 110 K was observed in the hexagonal FeGe [16-20]. As the first antiferromagnetic materials hosting CDW, understanding its correlation with magnetism has been one of the critical questions to be solved [21-23].

Compared to the cubic and the hexagonal phases, the monoclinic FeGe hosts the lowest symmetry, and its physical properties are much less reported. A previous study of the magnetic property on the monoclinic FeGe film was done by C. Zeng et al., and the films were grown on the Ge (111) substrates [24]. To reduce the interfacial intermixing, we synthesized the monoclinic FeGe films on $Al_2O_3$ (001). Structural characterizations show that the FeGe films are epitaxial to the substrate. These metallic films are ferromagnetic with a Curie temperature as high as ~ 800 K. At ~100 K, a doubled lattice periodicity was unambiguously captured under the transmission electron microscope (TEM).

We used pulse-laser deposition (PLD) to synthesize the monoclinic FeGe film. Before growth, the $Al_2O_3$ (001) substrates were sonicated in acetone and isopropyl alcohol, followed by a further degas at 400 °C until the pressure was lower than $1 \times 10^{-4}$ Pa. The advantage of the PLD is maintaining the elemental stoichiometry of the film as the target. Thereafter, the atomic ratio of the target (Fe: Ge) is designated as 1:1. During the growth, the substrate-target distance was kept at 5.0 cm and the fluence of the 248 nm excimer laser was 2 $J/cm^2$ with a repetition of 2 Hz. We systematically optimized the growth parameters and found that the FeGe films host the strongest X-ray intensities in the Bragg peaks when grown at 625 °C. The films' thicknesses (~ 35 nm) are determined by counting the laser pulse during the growth. The growth rate is ~ 2.0 Å per 100 pulses calibrated using X-ray reflection.

Like FeSn and Mn3Ge [25, 26], the FeGe films tend to be discontinuous and the two-probe resistances are too large to be measured by a multimeter. Here we adopt the 3-step growth technique: high temperature for the seed layer, low temperature for the continuous layer, and post-annealing for the pure phase [27]. For the seed layer, the growth took place at 625 °C for 1 hour; while for the continuous layer, the substrate was maintained at 550 °C for another 1 hour's growth. After a post-annealing at 625 °C for 30 minutes, the heater was turned off.

As shown in Fig. 1 (b), X-ray diffraction (Cu Kα1) shows that besides the $Al_2O_3$ (006) peak, two film peaks can be observed at *44.43°* and *98.19°* corresponding to lattice constants of *2.037 Å* and *1.019 Å* respectively. The cubic phase of FeGe can be easily ruled out. However, these two film peaks are close to both monoclinic (202) and (404) as well as hexagonal (002) and (004). Fortunately, there should be (001) and (003) peaks with measurable intensity for the hexagonal phase while neither (101) nor (303) peaks exist for the monoclinic phase. Our measurement shows no peaks at ~ *21.90°* or ~ *69.50°*, which suggests that these films are monoclinic. The measured (202) lattice constant is comparable to *2.047 Å* of bulk FeGe, and the slight variation may originate from the effect of substrate or possible defects in the film. The surface morphology of a FeGe film was characterized by an atomic force microscope (AFM). As shown in Fig. 1 (c), the surface is homogeneous with a root-meanF-square roughness of ~ 3.36 nm. Considering the giant structural difference, our growth of monoclinic FeGe on the trigonal $Al_2O_3$ substrate is surprising. To clarify the atomic coordination, the stacked $Al_2O_3$ (001) plane and FeGe (202) plane are plotted, as shown in Fig. 1 (d). For clarity, only O atoms are shown in the $Al_2O_3$ (001) plane. Regardless of their lattice constants, these two planes show similar geometry in most areas. However, there are discrepancies in certain areas as outlined in Fig. 1 (d). As a result, defects such as mosaicity may be introduced during the growth.

We further checked the in-plane coordination between the FeGe films and the $Al_2O_3$ substrate by performing azimuthal ϕ scans choosing the FeGe {112} and $Al_2O_3$ {014} planes. As shown in Fig. 2 (a), a threefold symmetry was captured in the $Al_2O_3$ when rotating along the *c*-axis. For the FeGe {112} planes, only two peaks were captured indicating a twofold symmetry, consistent with its monoclinic phase. We notice that the FeGe {112} peaks are broad in the ϕ scan. Considering the structural differences and the lattice mismatch between the film and the substrate, as discussed in Fig. 1 (d), this broadening is possibly due to the large mosaicity. To visualize its atomic structure directly, a high-resolution transmission electron microscope (HRTEM) was applied, and the specimen was fabricated using a typical focused-ion-beam technique. As shown in Fig. 2 (b), a sharp interface between FeGe and $Al_2O_3$ can be observed when viewing along [10-1] for FeGe and [-110] for $Al_2O_3$. A zoom-in area in FeGe shows that the atomic structure matches the crystal model well. After all, we confirm that the synthesized FeGe films are monoclinic with sharp interfaces between the $Al_2O_3$ substrates.

Transport measurement is an effective way of revealing materials' physical properties. In the cubic FeGe, the skyrmions' behaviors can be reflected by the topological Hall effect [1,2]. For the hexagonal FeGe, the formation of the CDW accompanies an increase of resistance in the transport measurement [16]. Here, we characterized the physical properties in the monoclinic FeGe films by transport. Before the measurements, the FeGe films were patterned into Hall bar geometry with ultra-violet lithography and Ar-ion milling. Both longitudinal and transverse resistances were measured. An optical image of the device is shown in the inset of Fig. 3 (a) where the scale bar is 200 μm. The film exhibits a metallic behavior with a slight upturn below 10 K. In general, the upturn at low temperatures can be induced by electron-electron interaction [28], weak localization [29], and the Kondo effect [30]. Detailed analysis of its origin is beyond the scope of this work. As shown in Fig. 3 (a), there are no observable kinks in the R-T curve or its first derivative. With the magnetic field applied perpendicular to the film, the magnetoresistances from 300 K to 10 K are shown in Fig. 3 (b). In general, the field-dependences of magnetoresistance behave similarly in the whole temperature range. When the field is lower than 1.5 T, quadratic and positive magnetoresistance up to 0.18% shows up. In the inset of Fig. 3 (b), the low-field magnetoresistance is hysteretic. The positive magnetoresistance at low fields exhibits orbit-effect-induced behavior. Further increase of external field changes the sign of the magnetoresistance which decreases to -0.68% at 9 T, indicating the formation of long-range ferromagnetic order.

Magnetic properties of the FeGe film were characterized by the magnetization and the Hall measurements, as shown in Fig. 3 (c) and Fig. 3 (d). During these measurements, the applied magnetic fields are perpendicular to the films. From 300 K to 10 K, the measured magnetization behaviors show no distinct differences except an increase in the saturation magnetization. Compared to the monoclinic FeGe bulk which is antiferromagnetic [31], our FeGe films are ferromagnetic. This is because the antiferromagnetism is destabilized due to the lattice compression, and the suppression of the magnetic moments is induced by increased *p-d* hybridization [24]. In our temperature-dependent measurement, its Curie temperature $T_c$ is as high as ~ 800 K [inset of Fig. 3 (c)]. In the field below 0.1 T, hysteric loops appear in all the temperature ranges we measured [inset of Fig. 3 (c)]. This may be related to short-range ordered local moments induced by defects. As shown in Fig. 3 (d), the anomalous Hall signals in the FeGe film behave similarly to the magnetization in Fig. 3 (c). By single-band model fitting of the high field data, we show that the carrier is n-type with density ranging from $1.08 \times 10^{23}$ cm$^{-3}$ at 300 K to $5.97 \times 10^{22}$ cm$^{-3}$ at 10 K [inset of Fig. 3 (d)].

In the hexagonal FeGe, there is a formation of CDW accompanied by lattice distortion, where an extra order in reciprocal space can be observed [16]. As mentioned above, the monoclinic and the hexagonal FeGe share similarities in structure (parts of their Bragg peaks are close). To check whether there are any structural transitions in the monoclinic FeGe film, we cooled the TEM specimen down to 100 K and monitored its crystal structure. Different from the room temperature HRTEM image of monoclinic FeGe in Fig. 2 (b), a stripe order emerges at 100 K as shown in Fig. 4 (a). Fig. 4 (b) and Fig. 4 (d) are zoom-in HRTEM images of FeGe structure at 300 K and 100 K, respectively. As can be seen, long stripes with a doubled periodicity orient along the out-of-plane direction of the film. From Fig. 4 (a), one can easily tell that the stripes nucleate near the interface and grow perpendicularly to the film. Compared to the left side, the right side of Fig. 4 (a) contains more disorders near the interface. Consequently, nucleation is easier for the right area where the stripes almost penetrate the film. Consistently, an extra set of peaks [such as (010)] appears at 100 K in the reciprocal space, Fig. 4 (c) and Fig. 4 (e). Compared to the original peaks such as (202) and (020), the full width at half maxima (FWHM) of the (010) peak is ~ 2 to 3 times broader, consistent with the inhomogeneity of the stripe order shown in Fig. 4 (a). Therefore, we confirm that similar to the hexagonal FeGe, a structural transition is also present in the monoclinic FeGe film near 100 K. In our measurements, the stripes are still partially formed, suggesting that 100 K is not low enough for a complete transition. Since the lowest accessible temperature is 100 K using our cryogenic sample holder, characterization of the full transition process at lower temperatures in the future is of great significance.

In summary, we have synthesized monoclinic FeGe films on Al$_2$O$_3$ (0001) and compared them with the cubic and the hexagonal phases. Magnetically, the monoclinic FeGe is ferromagnetic with Curie temperature $T_c$ as high as ~ 800 K. Structurally, we have observed a lattice distortion (~ 100 K) in the monoclinic FeGe film where stripes with doubled periodicity orient perpendicularly to the film. There are still open questions such as: at what temperature, will the lattice distortion fully complete? Is the lattice distortion related to CDW? To answer these questions, characterizations such as ARPES, Raman scattering, and DFT calculation are needed. Overall, our results enrich the FeGe phase diagram and suggest FeGe film as an ideal platform for investigation of the lattice structure, CDW, and magnetic order.

This work at Chongqing University is supported by the National Natural Science Foundation of China (Grant No. 12404123, 12374173, and 12374460), the Fundamental Research Funds for the Central Universities (Grant No. 2023CDJXY-0049). We thank Dr. Dashuai Ma, Dr. Yan Liu, and the Testing Center of Chongqing University for the fruitful discussion, and assistance in transport measurement, magnetic property measurement, and structural characterizations.

**Reference:**

[1] S. X. Huang, and C. L. Chien, Phys. Rev. Lett. 108, 267201 (2012).

[2] J. C. Gallagher, K. Y. Meng, J. T. brangham, H. L. Wang, B. D. Esser, D. W. McComb, and F. Y. Yang, Phys. Rev. Lett. 118, 027201 (2017).

[3] K. Shibata, A. Kovács, N. S. Kiselev, N. Kanazawa, R. E. Dunin-Borkowski, and Y. Tokura, Phys. Rev. Lett. 118, 087202 (2017).

[4] X. Z. Yu, N. Kanazawa, Y. Onose, K. Kimoto, W. Z. Zhang, S. Ishiwata, Y. Matsui & Y. Tokura, Nat. Mater. 10, 106-109 (2011).

[5] T. Ohoyama, K. Kanematsu and K. Yasukochi, J. Phys. Soc. Japan 18, 589 (1963).

[6] J. Bernhard, B. Lebech and O. Beckman, J. Phys. F: Met. Phys. 14, 2379 (1984).

[7] J. Bernhard, B. Lebech and O. Beckman, J. Phys. F: Met. Phys. 18, 539 (1988).

[8] H. -M. Guo, and M. Franz, Phys. Rev. B 80, 113102 (2009).

[9] W. Beugleling, J. C. Everts, and C. M. Smith, Phys. Rev. B 86, 195129 (2012).

[10] Z. Li, J. Zhuang, L. Wang, H. Feng, Q. Gao, X. Xu, W. Hao, X. Wang, C. Zhang, K. Wu, S. X. Dou, L. Chen, Z. Hu, Y. Du, Sci. Adv. 4, eaau4511 (2018).

[11] W. -S. Wang, Z. -Z. Li, Y. -Y. Xiang, and Q. -H. Wang, Phys. Rev. B 87, 115135 (2013).

[12] E. Tang, J.-W. Mei, and X.-G. Wen, Phys. Rev. Lett. 106, 236802 (2011).

[13] T.-H. Han, J. S. Helton, S. Chu, D. G. Nocera, H. A. Rodriguez-Rivera, C. Broholm & Y. S. Lee, Nature 492, 406-410 (2012).

[14] B. R. Oritz, S. M. L. Teicher, Y. Hu, J. L. Zuo, P. M. Sarte, E. C. Schueller, A. M. M. Abeykkoon, M. J. Krogstad, S. Rosenkranz, R. Osborn, R. Seshadri, L. Balents, J. He, and S. D. Wilson, Phys. Rev. Lett. 125, 247002 (2020).

[15] S. Yan, D. A. Huse, S. R. White, Science 332, 1173-1176 (2011).

[16] X. Teng, L. Chen, F. Ye, E. Rosenberg, Z. Liu, J.-X. Yin, Y.-X. Jiang, J. S. Oh, M. Z. Hasan, K. J. Neubauer, B. Gao, Y. Xie, M. Hashimoto, D. Lu, C. Jozwiak, A. Bostwick, E. Rotenberg, R. J. Birgeneau, J.-H. Chu, M. Yi & P. Dai, Nature 609, 490-495 (2022).

[17] S. Shao, J.-X. Yin, I. Belopolski, J.-Y. You, T. Hou, H. Chen, Y. Jiang, M. S. Hossain, M. Yahyavi, C.-H. Hsu, Y. P. Feng, A. Bansil, M. Z. Hasan, and G. Chang, ACS Nano 17, 10164-10171 (2023).

[18] H. Miao, T. T. Zhang, H. X. Li, G. Fabbris, A. H. Said, R. Tartaglia, T. Yilmaz, E. Vescovo, J.-X. Yin, S. Murakami, X. L. Feng, K. Jiang, X. L. Wu, A. F. Wang, S. Okamoto, Y. L. Wang & H. N. Lee, Nat. Commun. 14, 6183 (2023).

[19] L. Chen, X. Teng, H. Tan, B. L. Winn, G. E. Granroth, F. Ye, D. H. Yu, R. A. Mole, B. Gao, B. Yan, M. Yi, & P. Dai, Nat. Commun. 15, 1918 (2024).

[20] Z. Zhao, T. Li, P. Li, X. Wu, J. Yao, Z. Chen, S. Cui, Z. Sun, Y. Yang, Z. Jiang, Z. Liu, A. Louat, T. Kim, C. Cacho, A. Wang, Y. Wang, D. Shen, J. Jiang, and D. Feng, arXiv: 2308.08336.

[21] B. Zhang, J. Ji, C. Xu, H. Xiang, Phys. Rev. B 110, 125139 (2024).

[22] C. Shi, Y. Liu, B. B. Maity, Q. Wang, S. R. Kotla, S. Ramakrishnan, C. Eisele, H. Agarwal, L. Noohinejad, Q. Tao, B. Kang, Z. Lou, X. Yang, Y. Qi, X. Lin, Z.-A. Xu, A. Thamizhavel, G.-H. Cao, S. van Smaalen, S. Cao, and J.-K. Bao, Sci. China Phys. Mech. Astron. 67, 117012 (2024).

[23] X. Wu, X. Mi, L. Zhang, X. Zhou, M. He, Y. Chai, A. Wang, Phys. Rev. Lett. 132, 256501 (2024).

[24] C. Zeng, P. R. C. Kent, M. Varela, M. Eisenbach, G. M. Stocks, M. Torija, J. Shen, and H. H. Weitering, Phys. Rev. Lett. 96, 127201 (2006).

[25] D. Hong, C. Liu, H.-W. Hsiao, D. Jin, J. E. Pearson, J.-M. Zuo, and A. Bhattacharya, AIP Adv. 10, 105017 (2020).

[26] A. Markou, J. M. Taylor, A. Kalache, P. Werner, S. S. P. Parkin, and C. Felser, Phys. Rev. Mater. 2, 051001(R) (2018).

[27] D. Hong, C. Liu, J. Wen, Q. Du, B. Fisher, J. S. Jiang, J. E. Pearson, and A. Bhattacharya, APL Mater. 10, 101113 (2022).

[28] S. Tong, X. Zhao, D. Wei, and J. Zhao, Phys. Rev. B 101, 184434 (2020).

[29] A. Bagri, S. Sahoo, R. J. Choudhary, D. M. Phase, J. Alloys Compd. 902, 163644 (2022).

[30] Y. Wang, C. Xie, J. Li, Z. Du, L. Cao, Y. Han, L. Zu, H. Zhang, H. Zhu, X. Zhang, Y. Xiong, and W. Zhao, Phys. Rev. B 103, 174418 (2021).

[31] G.P. Felcher and J. D. Jorgensen, J. Phys. C 16, 6281 (1983).

Fig. 1

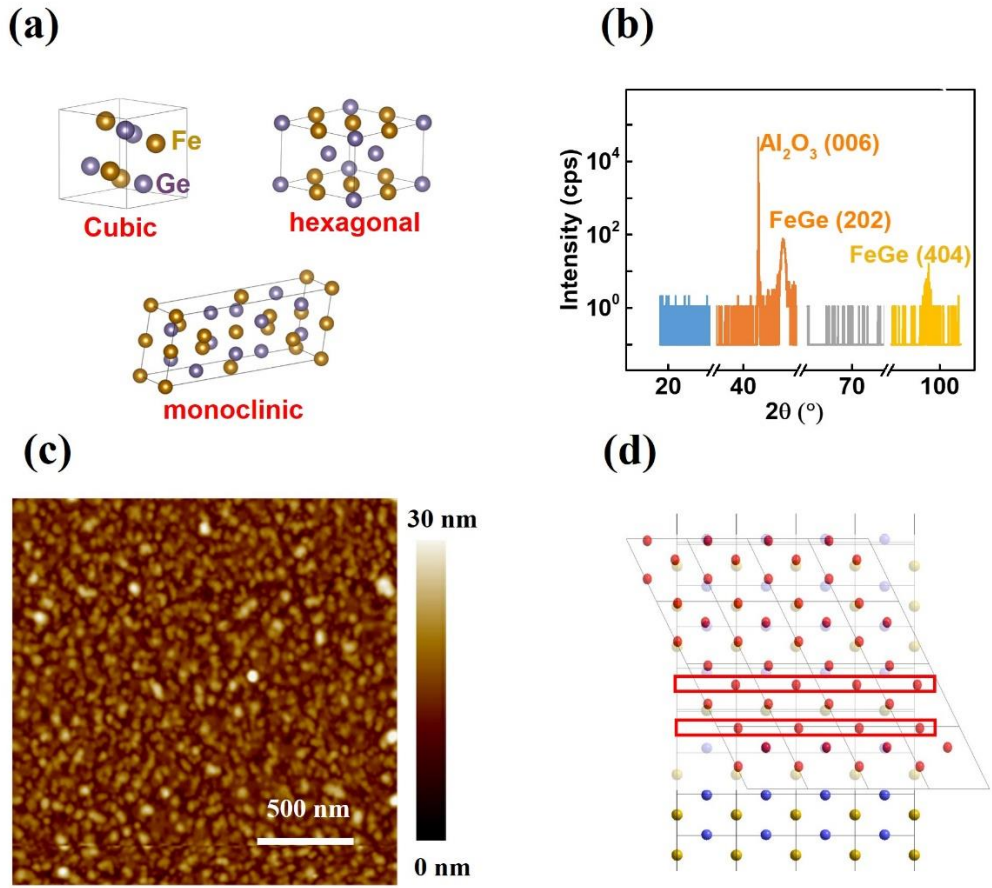

Fig.1. (a) Crystal structures of the cubic, hexagonal, and monoclinic FeGe. (b) X-ray diffraction of FeGe film on $Al_2O_3$ (001). Breaks are added for clarity. (c) AFM image of a FeGe film. (d) Stacking of FeGe (202) and $Al_2O_3$ (001) planes. Only O atoms (red) are shown in the $Al_2O_3$ plane.

Fig. 2

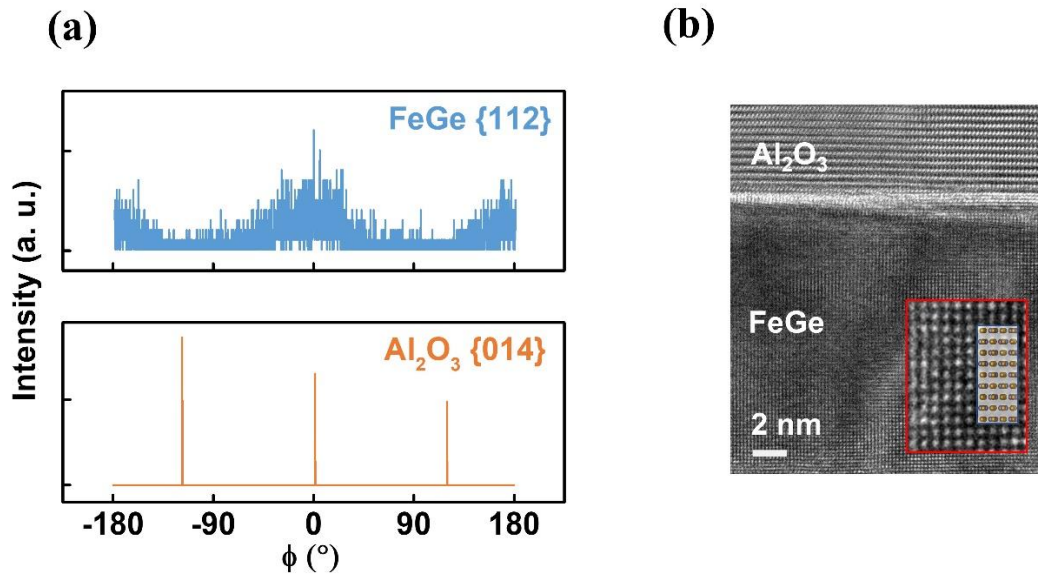

Fig.2. (a) Phi-scans around the in-plane peaks (as labeled) of FeGe and $Al_2O_3$. (b) Cross section of the FeGe/$Al_2O_3$ at 300 K and the viewing direction is along $Al_2O_3$ [-110]. Inset shows the comparison between a zoom-in of the measured lattice structure and the crystal model.

**Fig. 3**

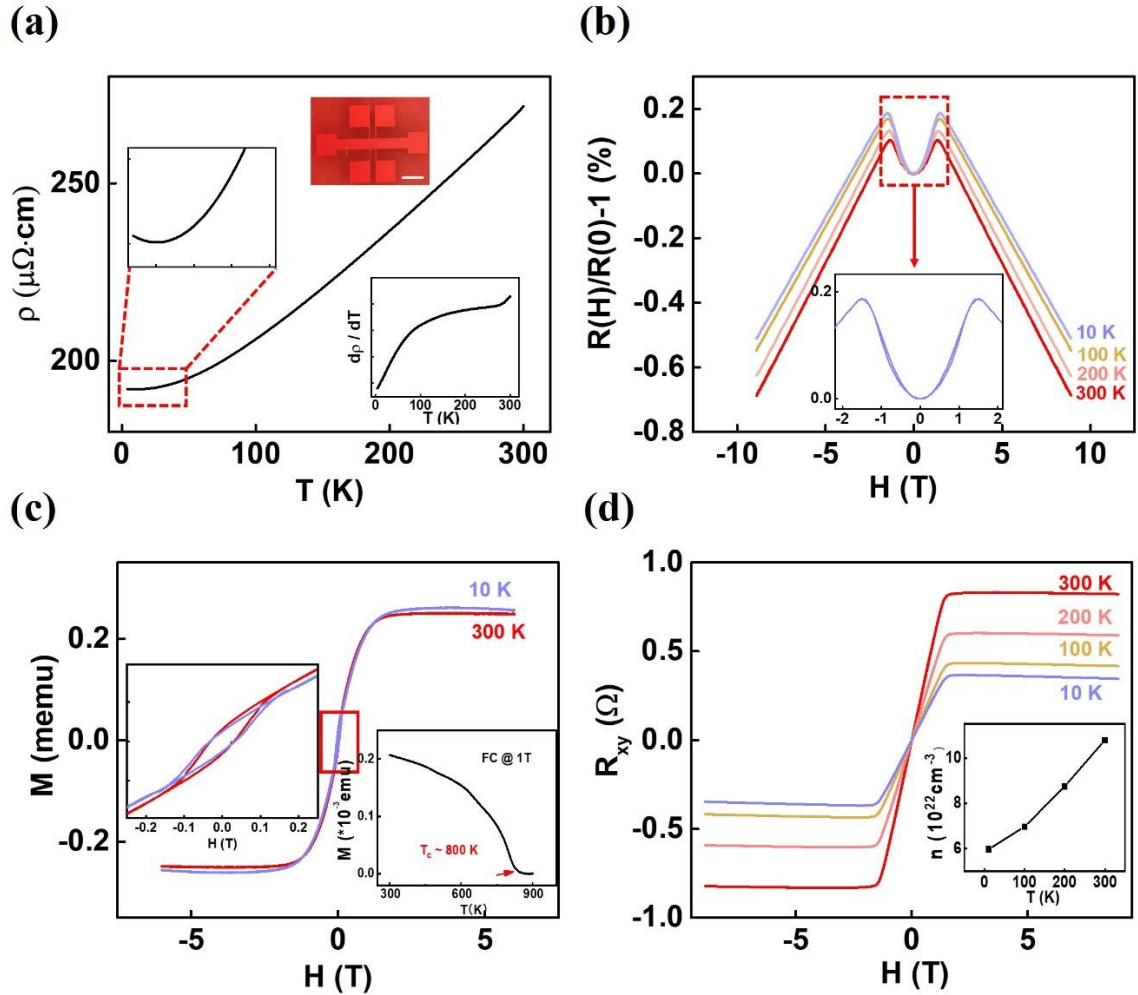

Fig.3. (a) Temperature dependence of resistivity in FeGe/Al$_2$O$_3$ (001) (35 nm thick). Insets: zoom-in area at low temperatures, an optical image of the patterned device (scale bar ranges 200 μm), and its first derivative of the measured ρ-T curve. (b) Magnetoresistance of FeGe/Al$_2$O$_3$ (001), and the low-field hysteresis is shown in the inset. (c) Magnetization curve at 300 K and 10 K with external field perpendicular to the film (5 mm x 5 mm x 35 nm), and the insets show the temperature dependence of magnetization in 1 T field (right) and the low-field hysteresis (left). (d) Anomalous Hall effect in FeGe/Al$_2$O$_3$ (001). Inset: calculated carrier density by analyzing high-field data using a single-band model.

**Fig. 4**

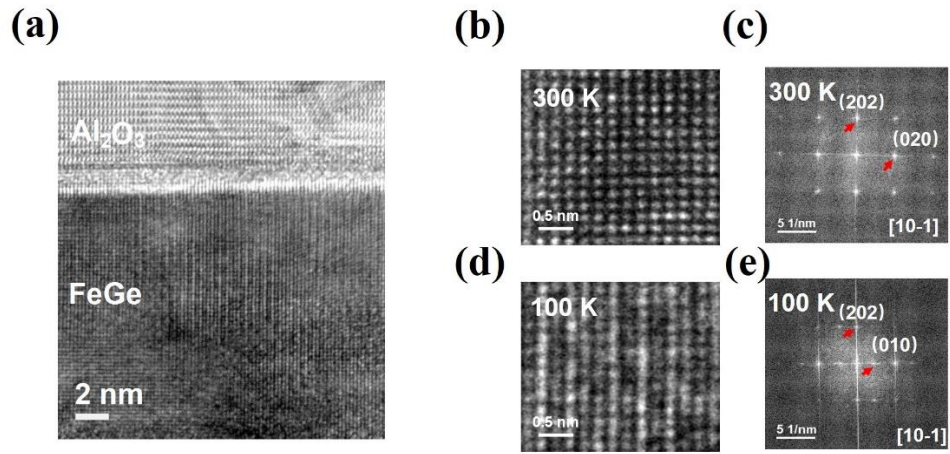

Fig.4. (a) Cross section of FeGe/Al$_2$O$_3$ at 100 K viewing along Al$_2$O$_3$ [-110]. (b) and (d) are zoom-in areas for FeGe at 300 K and 100 K respectively. (c) and (e) are the reciprocal images of (b) and (d), respectively.